\documentclass[useAMS,usenatbib,psfig]{mn2e}
\usepackage{graphicx}
                                   
\def\ang{\mbox{\AA}}

\def\kms{\mbox{${\rm kms}^{-1}$}}

\def\asec{\hbox{$^{"}$}}

\def\etal   {et~al.\ }

\def\name#1 {{\it #1\/}}
\def\vol#1  {{\bf #1\/}}
\title[Isolated Ellipitical Kinematics]{Radial Kinematics of Isolated Elliptical Galaxies}
\author[Hau \& Forbes]{G.K.T. Hau$^{1}$\thanks{E-mail:
ghau@eso.org}, Duncan A. Forbes$^{2}$\\
$^{1}$Department of Physics, University of Durham, South Road,
Durham, DH1 3LE\\
$^{2}$Centre for Astrophysics and Supercomputing, Swinburne University of Technology, Hawthorn, VIC 3122, Australia}

\begin{document}
\date{\today}

\pagerange{\pageref{firstpage}--\pageref{lastpage}} \pubyear{2004}

\def\zsun{\mbox{${\rm Z}_{\odot}$}}
\def\square{\vrule height 4.5pt width 4pt depth -0.5pt}
\def\threesquares{\square~\square~\square\ }
\def\remark#1{{\threesquares\tt#1~\threesquares}}
\newcommand{\mean}[1]{\mbox{$<$#1$>$}}
\newcommand{\h}[1]{\mbox{$h_#1$}}
\newcommand{\hbeta}{\mbox{H${\beta}$}}
\newcommand{\mg}[1]{\mbox{Mg$_#1$}}
\newcommand{\mgb}{\mbox{Mg$\ b$}}
\newcommand{\fe}[1]{\mbox{Fe$_{#1}$}}

\maketitle
\begin{abstract}

Ellipticals in very low density environments are extremely rare
but hold important clues about galaxy formation and evolution. In
this paper we continue our study of isolated elliptical galaxies,
presenting results on the radial stellar kinematics for 13
isolated early-type galaxies. 
We derive radial rotation velocity, velocity dispersion and hermite terms
to $\sim$ 1 effective radius. We 
observe a dichotomy in kinematic properties similar to that in the
elliptical population as a whole, where low luminosity
ellipticals tend to be rotationally supported.
For all galaxies
the V/$\sigma$ ratio increases with radius. 
We find kinematically distinct cores (KDCs), or velocity
substructure, in $\sim 40\%$ of the galaxies for which we have
major axis spectra. 
Such a fraction is similar to that observed
for ellipticals in higher density environments.  
Most galaxies in the sample reveal kinematic evidence for a
nuclear disk. The non-relaxed
kinematics in several galaxies suggests that they have undergone 
a merger or accretion event.
Isolated ellipticals generally follow the
fundamental plane defined by cluster ellipticals -- exceptions
being those galaxies with evidence for young stellar
populations. Overall, we find isolated ellipticals have similar kinematic
properties to their counterparts in higher density environments.

\end{abstract}

\begin{keywords}
galaxies: elliptical and lenticular, galaxies: general -
galaxies: kinematics and dynamics, galaxies: fundamental parameters 
\end{keywords}


\section{Introduction} \label{sec:introduction}

Galaxies are found in three broad environments, defined in terms
of their spatial density. These range from clusters with hundreds
of galaxies within a few Mpc, to groups to relatively isolated
galaxies (the `field'). The local environment is thought to play a key role
in galaxy formation and subsequent evolution.

It is relatively easy to study large numbers of cluster galaxies
because a single CCD image or spectral mask may contain hundreds of
galaxies. For low density environments, the same CCD may contain only
one or two bright galaxies.  Consequently, our knowledge of the nearby
Universe is often directly proportional to the environmental
density. Very isolated galaxies (sometimes called the `extreme field')
are consequently the least studied and understood.

Although poorly studied, this extreme in the density range of galaxy
environments offers crucial `leverage' to tackle several outstanding
issues in galaxy evolution.  For example, the current hierarchical
models for galaxy formation predict that ellipticals in low density
environments have younger mean stellar populations and lower [Mg/Fe]
ratios than their cluster counterparts (e.g.  Kauffmann \& Charlot
1998; Kuntschner et al. 2002; Nagashima et al. 2005; de Lucia
et al. 2005). Under this framework, isolated ellipticals are thought to
have formed in relatively recent mergers of spiral galaxy pairs. If
so, we might expect a high incidence of kinematical misalignments
(Weil \& Hernquist 1996), such as kinematically distinct cores
(KDCs). Alternatively, large isolated ellipticals may be the result of
a merged small group of galaxies (Jones, Ponman \& Forbes 2000;
D'Onghia et al. 2005). Here we might expect an increasing importance
of rotation with galactocentric radius and rotation about the minor
axis (Weil \& Hernquist 1996).

Kuntschner et al. (2002) studied a sample of nine early-type galaxies,
selected from Southern Sky Survey plates, in low density environments
(i.e. $\le$ 2 neighbours within 1.3 Mpc). Their CCD imaging revealed
most of the galaxies to be morphologically disturbed; six of the nine
are listed in the Arp-Madore catalogue of peculiar galaxies. They
obtained long-slit spectra of the galaxies at the MSSSO 2.3m
telescope. They concluded that their sample galaxies were 2--3 Gyrs
younger and 0.2 dex more metal-rich than E/S0 galaxies in the Fornax
cluster. The [Mg/Fe] ratio was similar to the Fornax galaxies. They
did not explore the internal kinematics of their sample.  The imaging
study of 9 isolated early-type galaxies by Marcum et al. (2004) found
a range of galaxy colours and morphologies -- with some galaxies
suggestive of being a recent merger remnant, while others appeared to
be undisturbed and pristine.

In Reda et al. (2004) we defined a new sample of isolated
galaxies. Briefly, early-type galaxies outside of groups and clusters,
with B$_T$ $\le$ 14, were selected from the LEDA database. These were
further restricted to have no neighbours within 700 km/s in velocity,
0.67 Mpc on the sky and 2 B-band magnitudes. Finally, a visual check
using the DSS was carried out. This resulted in a sample of 36 very
isolated E/S0 galaxies. Wide-field imaging of this sample is 
reported in Reda et al. (2004). Like Kuntschner et al. (2002) and
Marcum et al. (2004), Reda et
al. found several isolated galaxies to have signs of recent
morphological disturbance.

In a second paper, Reda, Forbes \& Hau (2005) examined the location of
isolated galaxies on the fundamental plane (FP) using their
photometric properties and central velocity dispersions.  Most the
sample were found to lie on the FP as defined by cluster
ellipticals. However, a small number of morphologically disturbed
isolated galaxies deviated strongly from the FP in a direction
consistent with their younger stellar populations (e.g. Forbes, Ponman
\& Brown 1998).

Here we present new long-slit spectra of 13 isolated
galaxies from the sample of Reda et al. (2004) and investigate their radial kinematic properties. 
A future paper will explore radial stellar population trends for
these galaxies.


\section{Observations and data reduction} \label{sec:observations}

\subsection{Observations}

A subsample of 13 isolated ellipticals from Reda et al. (2004) were
observed over two observing runs (i.e. 2002 Jan. 16--18 and 2004
Dec. 11--12) using EFOSC2 at the ESO 3.6m on the La Silla
Observatory. The observational setup for the runs are given in
Table 1. The exposure times were $2 \times 1200$ seconds for the 2002
run, and $3 \times 1200$ seconds for the 2004 run.

For the elliptical galaxies we generally placed the slit along the major
axis, but for the S0 galaxy ESO 153-G003 we chose to avoid the disk and hence
aligned the slit along the minor axis. For NGC 1162 where no preferred axis was
obvious, we choose the parallactic angle. The observed position angle
(P.A.) of the slit is given, along with properties of the galaxy sample in
Table 2.

Lick and spectrophotometric standards stars were taken at the
parallactic angle. The Lick stars are also used for velocity
standards.  The spectrophotometric standards were taken with a 5$^{''}$
wide slit.

\begin{table}
\begin{center}
\begin{tabular}{lc}
\hline
Run 1 & 2002\\
\hline
Chip &  Loral/Lesser \\
Chip size (pixels)& 2048$\times$2048 \\
Binning & $2 \times 2 $ \\
Pixel scale (binned pixel) & 0.314\asec \\ 
Read noise (e$^-$/binned pixel) &  9.2 \\
Gain (e$^-$/ADU) & 1.33 \\
Grism\# & 8 \\
Lines/mm & 600\\
Grism resolution FWHM & 9.3\AA\\
Wavelength coverage & 4320 -- 6360\AA \\
Slit width & 1.5 \asec \\
\hline
\end{tabular}
\begin{tabular}{lc}
Run 2 & 2004\\
\hline
Chip &  Loral/Lesser \\
Chip size (pixels)& 2048$\times$2048 \\
Binning & $2 \times 2 $ \\
Pixel scale (binned pixel) & 0.314\asec \\ 
Read noise (e$^-$/binned pixel) &  10.3 \\
Gain (e$^-$/ADU) & 1.17 \\
Grism\# & 8 \\
Lines/mm & 600\\
Grism resolution FWHM & 7.8\AA\\
Wavelength coverage & 4320 -- 6360\AA \\
Slit width & 1.2 \asec \\
\hline
\end{tabular}
\caption{EFOSC2 observing characteristics.}
\label{'CCD'}
\end{center}
\end{table}

\begin{table*}
\begin{center}
\begin{tabular}{lrrcccccccccc}
\hline 
Galaxy & R.A. & Dec. & T & $V_{Virgo}$& $M_B$ & $A_B$ &
R$_e$ & $<\mu_e>$ & Slit P.A. & Gal P.A. & Ellip. & Run \\
       & (J2000) & (J2000) &        & (km/s) & (mag) &
(mag)        & (\asec) & (mag/sq.'') & (deg) & (deg)& & \\
\hline 
ESO 153-G003 & 01 58 18.2 & -54 12 57 &-3.9&  6242 &-20.12&
0.12& 10.2 & 20.6 & 291	& 21 & 0.5 & 2\\	
ESO 218-G002 & 12 21 09.3 & -52 35 08 &-3.9&  4072 &-20.62&
0.75& 20.1 & 21.5 & 187   & 3 & 0.15 & 1 \\ 
ESO 318-G021 & 10 53 05.2 & -40 19 46 &-4  &  4641 &-20.71&
0.35& 21.7 & 21.7 & 301 & 121 & 0.3 & 1\\
NGC 682 & 01 49 4.6 & -14 58 29 &-2.9&  5456 &-19.95& 0.07&
-- & -- & 321	& 96 & 0.3 & 2\\
NGC 1045 & 02 40 29.1 & -11 16 39 &-3.0& 4496 &-20.63& 0.19&
16.7 & 20.8 & 238 & 58 & 0.3 & 2\\
NGC 1162 & 02 58 56.0 & -12 23 55 &-5.0& 3789 &-20.22& 0.20&
-- & -- & 280 & 51 & 0.1 & 2\\	 	
NGC 2110 & 05 52 11.4 & -07 27 24       &-3  &  2065 &-19.82&
1.62 & 24.3 & 20.7 & 345 & 345  & 0.2 & 1 \\
NGC 2271 & 06 42 52.2 & -23 28 32      &-3.2&  2402 &-19.86&
0.53& 19.0 & 20.9 & 269  & 84 & 0.3 & 1\\
NGC 2865 & 09 23 30.7 & -23 09 48      &-4.1&  2472 &-20.35&
0.36& 27.7 & 20.8 & 333 & 153  & 0.2 & 1\\
NGC 4240 & 12 17 24.3 & -09 57 07      &-4 &   1916 &-18.87&
0.23&   -- & -- & 276  & 103  & 0.1 & 1\\
MCG-01-27-013 & 10 32 12.3 & -9 14 22 &-3 &   9044 &-20.90&
0.19&   13.8 & 22.0 & 191   & 11  & 0.45 & 1\\
MCG-02-13-009 & 04 48 12.7 & -13 40 03 &-4 &   5494 &-20.74&
0.42&   -- & -- & 329  & 149  & 0.1 & 1\\
MCG-03-26-030 & 10 11 19.0 & -17 12 17&-3 &   8948 &-21.50&
0.22&  15.6 & 21.4 & 284 & 104  & 0.25 & 1\\
\hline
\end{tabular}
\caption{The observed sample. The Hubble T-type, Virgo infall
  and corrected velocity $V_{Virgo}$ are from Hyperleda.  The B-band
  absolute magnitude $M_B$ is calculated from the apparent
magnitude and foreground extinction $A_B$ from Hyperleda, and 
  and velocity assuming $H_0 = 75$ \kms Mpc$^{-1}$. The B-band surface
  brightness is from Reda, Forbes \& Hau (2005). 
  The position angles of the slits are
  defined such that distances along the slit increase in the
  direction of the P.A. on the sky. The galaxy position angle is
  estimated from DSS images. The ellipticity is taken either from Reda
  et al. (2004, 2005) or from Hyperleda. Runs 1 and 2 correspond to
  2002 and 2004 respectively. }
\label{tab:galaxies}
\end{center}
\end{table*}

\subsection{Data reduction}

The data has been reduced using tasks within IRAF. After
bias-subtraction, the pixel-to-pixel response variation was corrected
using dome flats.  The wavelength calibration is good to within 0.5
\ang ~or 29 km/s rms. The data was corrected for instrumental response
and spatial distortion and rebinned to logarithmic wavelength for the
kinematic analysis. Special attention was taken with the slit-transfer
function. This was estimated using dome flats which give a more
uniform background than the twilight flats which are more prone to
scattered light due to the open nature of the telescope. The sky
background was determined from regions of the two-dimensional spectrum
far from the galaxy centre and subtracted from the final data.

\section{Kinematic analysis}

Kinematic analysis was performed on logarithmically-binned data, using
van der Marel's code (van der Marel 1994). This code first finds the
best fitting mean velocity and dispersion for a Gaussian LOSVD, and
then calculates the higher order moments such as the 3rd and 4th order
Hermite terms (\h3 and \h4). These terms measure asymmetric and
symmetric deviations from a Gaussian profile respectively (van der
Marel \& Franx 1993). Note that although we present the \h3 and \h4
terms, the velocity and velocity dispersions are recovered independent
of them.

Possible emission lines of [OII], [OIII] and [NI] (even if no emission
line is visible), the H$\beta$ and H$\gamma$ lines 
and any pixels where the sky is strong are masked
from the fitting. To arrive
at the final values, the pixels which are 4--5 sigma from the mean in
the residual spectrum are rejected, and this process is reiterated
until a stable solution is obtained. This helps to reject hidden
emission lines and bad columns or pixels.

\subsection{Template matching}

Template matching is important for the recovered kinematics as the
line-profiles can be sensitive to the choice of template. A number of
stars have been observed with the same instrumental setup as the
galaxy observations. These are: HD15772 (K2III), HD36003 (K5V), 
HR2429 (K1III), HR2574 (K4III), HR3428 (G9III), HR4287 (K1III) and
HR4657 (F5V).  The latter three are also Lick standards.  Template
mismatch is minimised by comparing the $\chi^2$ values from the LOSVD
fitting using different stars.  For each galaxy, a spectrum with S/N
of 100--140 which encompasses a region of ~ 1.6--12$^{``}$, depending
on the galaxy, was extracted and compared with each template in turn.
The best template with the lowest average $\chi^2$ is chosen. HR 2574
gave the best fits for all galaxies, and is used for the final data
analysis.
Further optimisation was done by an iterative process to reject pixels
which deviate too much from the fitted model. 

\subsection{Stellar dispersion measurements}

%

We measure velocity dispersions according to the prescription
of Jorgensen et al. (1995). They used an aperture
correction based on empirical kinematic models. Below we quote 
the equation which gives a distance-independent central velocity
dispersion by correcting for the fact that an aperture of fixed
angular size samples a larger portion of a radial velocity profile within
a more distant galaxy.    
\begin{equation}
log \frac{\sigma_{ap}}{\sigma_{norm}} = -0.04\ log \frac{r_{ap}}{r_{norm}}
\end{equation}
Here $\sigma_{ap}$ is the velocity dispersion measured in a circular aperture of radius
$r_{ap}$, the `normalised' velocity dispersion $\sigma_{norm}$ is
measured in the normalisation radius $r_{norm}$, i.e.
$2 r_{norm} = 3.4 \asec$ for a galaxy at the distance of
Coma. 

Using a library of kinematic data, they modelled the correction
for $\sigma_{ap}$ measured in a rectangular aperture of size $x \times y$
to that measured in a circular aperture of radius $r_{ap}$.
\begin{equation}
2 r_{ap} \approx 1.025 \times 2 (xy/\pi)^{1/2}
\end{equation}
We chose
to measure the sigma inside a rectangular aperture of size $1.5^{''}
\times 5^{''}$ for the 2002 data, and $1.2^{''} \times 5^{''}$ for the
2004 data, inside which the spectra are summed. The length of 5$^{''}$
along the slit is chosen such that it is at least 3 times the measured
seeing. The corresponding $r_{ap}$ is $1.584^{''}$. For the $1.2^{''}
\times 5^{''}$ aperture, the $r_{ap}$ is $1.417^{''}$.

Using equations 1 and 3 we can then relate the dispersion measured
in a square aperture to the normalised dispersion. 
For each spectrum, we
apply the kinematic analysis as described above.
We choose to fit over the spectrum from
$\lambda= 4420 \ang$ to $6280\ang$. The regions corresponding to within 3
$\sigma$ of an emission line or the H$\beta$ and H$\gamma$ lines are
excluded (even if there is no emission). At the end of each fit, the
standard deviation in the noise-weighted residuals is calculated.  If
a pixel has a value more than 3 standard deviations from the fitted
model, it is excluded from the next fit. The process is repeated
until a stable solution is obtained.  
The resulting $\sigma_{\rm{norm}}$ values and S/N ratios  
are listed in Table 4.


\section{Kinematics of individual galaxies} \label{sec:results}

Before presenting our kinematic profiles we briefly discuss the
interpretation of the \h3 and \h4 terms (see van der Marel \&
Franx 1993 for further details). As mentioned earlier, the
hermite terms quantify deviations from a simple Gaussian velocity
distribution. If the galaxy has a large-scale kinematic
coherence then we expect symmetry (or anti-symmetry) about the
galaxy centre. Simple equilibrium models may then be applicable
for the galaxy's kinematics. For the \h4
term we expect the profile to be similar on both sides of galaxy,
whereas for \h3 we expect a reflection about an angle of 45$^o$
i.e. positive one side and negative the other side of the galaxy
centre. Embedded disks can be detected in the kinematic profiles;
expected properties include a smooth anti-asymmetric velocity
profile with flat or declining velocity dispersion with radius,
anti-symmetric \h3 profile and weak \h4 term. 
Small systematic offsets are often seen in the \h4
term. This is due to the sensitivity of this term to continuum
subtraction. The profiles in the central galaxy regions can be
smeared out due to seeing effects, so any dips should be
interpreted with caution. 

In Figures 1--4 we show the velocity, velocity dispersion, \h3 and
\h4 radial profiles along the slit direction. 
Below we briefly
discuss these kinematic radial profiles for each galaxy and
summarise galaxy information from our previous imaging studies (Reda et
al. 2004, 2005) and from NED. 

\begin{figure*}
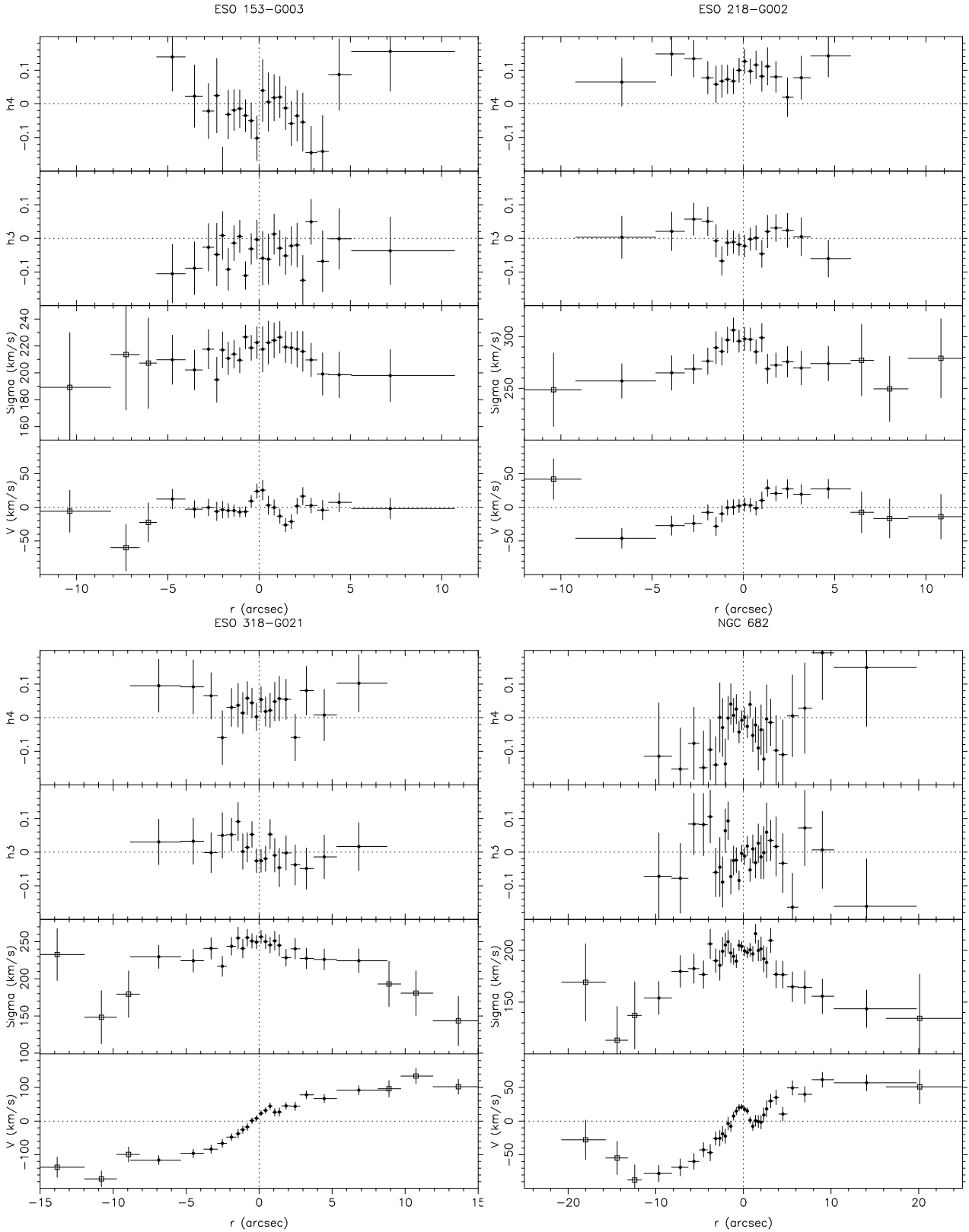

\begin{center}
\includegraphics[scale=0.42]{ESO153vsh3h4.ps}
\includegraphics[scale=0.42]{ESO218vsh3h4.ps}
\includegraphics[scale=0.42]{ESO318vsh3h4.ps}
\includegraphics[scale=0.42]{N682vsh3h4.ps}
\caption{Radial kinematic profiles of isolated galaxies. From
bottom to top are shown rotation velocity, velocity dispersion,
3rd and 4th hermite terms. The open symbols indicate lower S/N
data for which we do not calculate hermite terms.}
\end{center}
\end{figure*}

\begin{figure*}
\begin{center}
\includegraphics[scale=0.42]{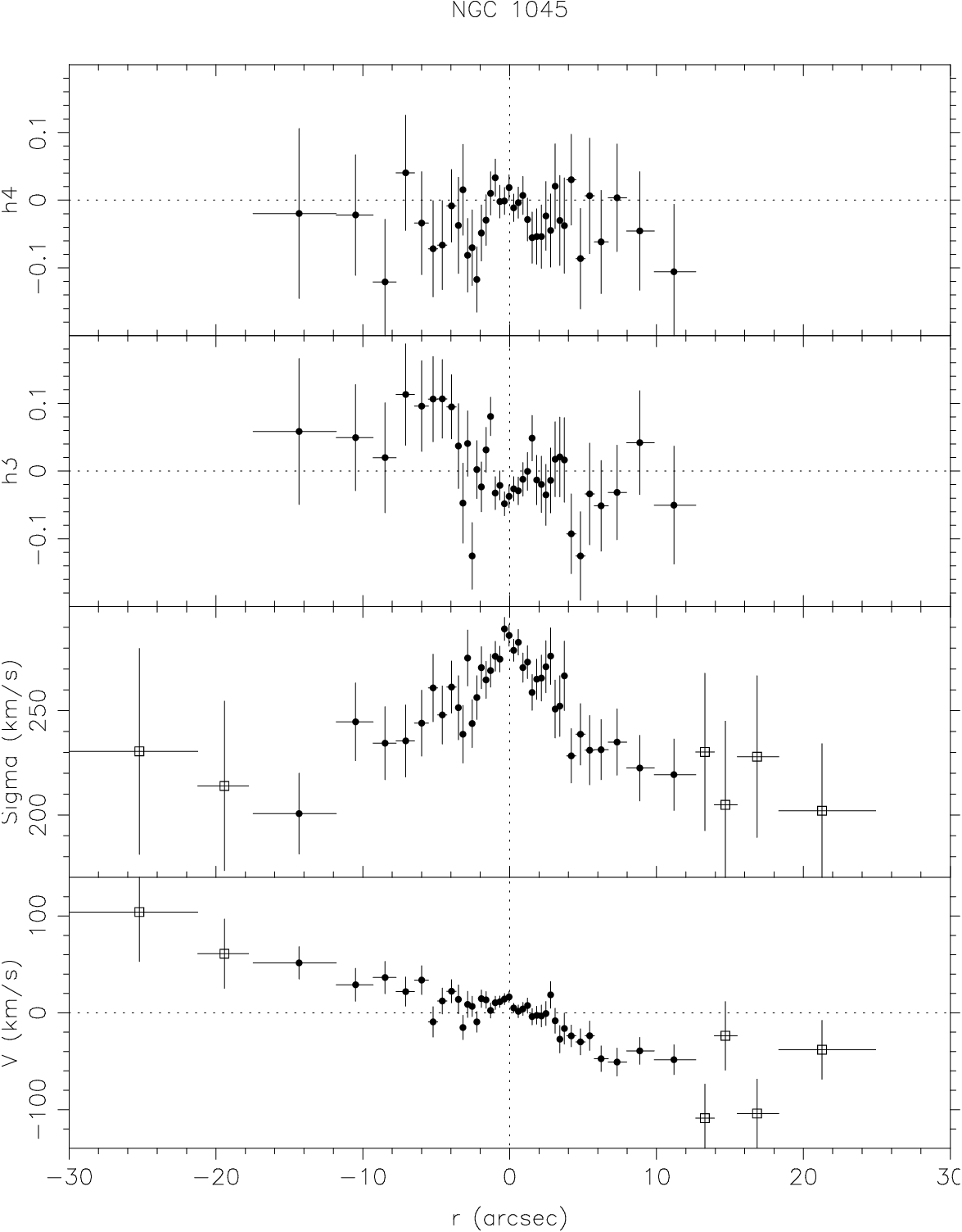}
\includegraphics[scale=0.42]{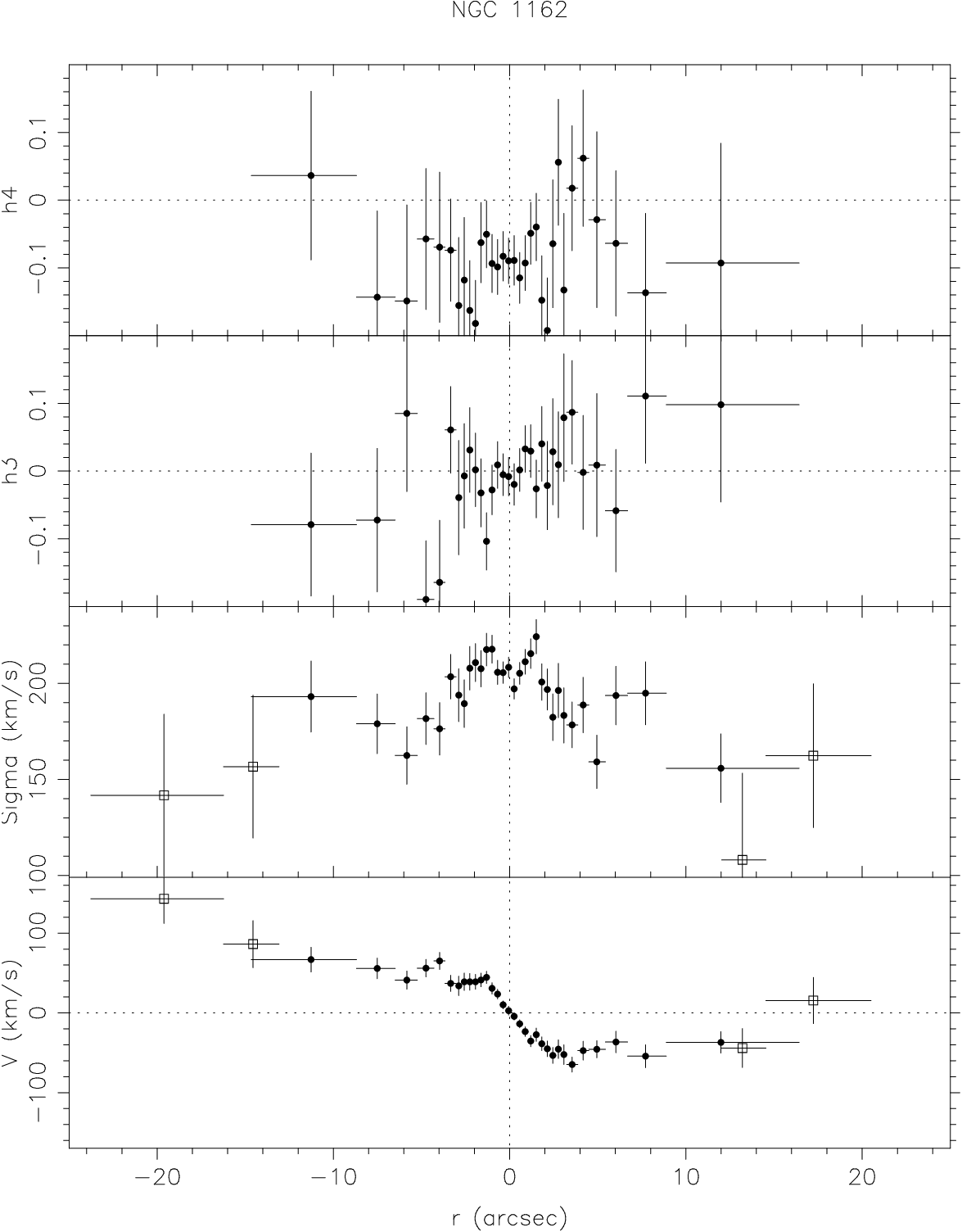}
\includegraphics[scale=0.42]{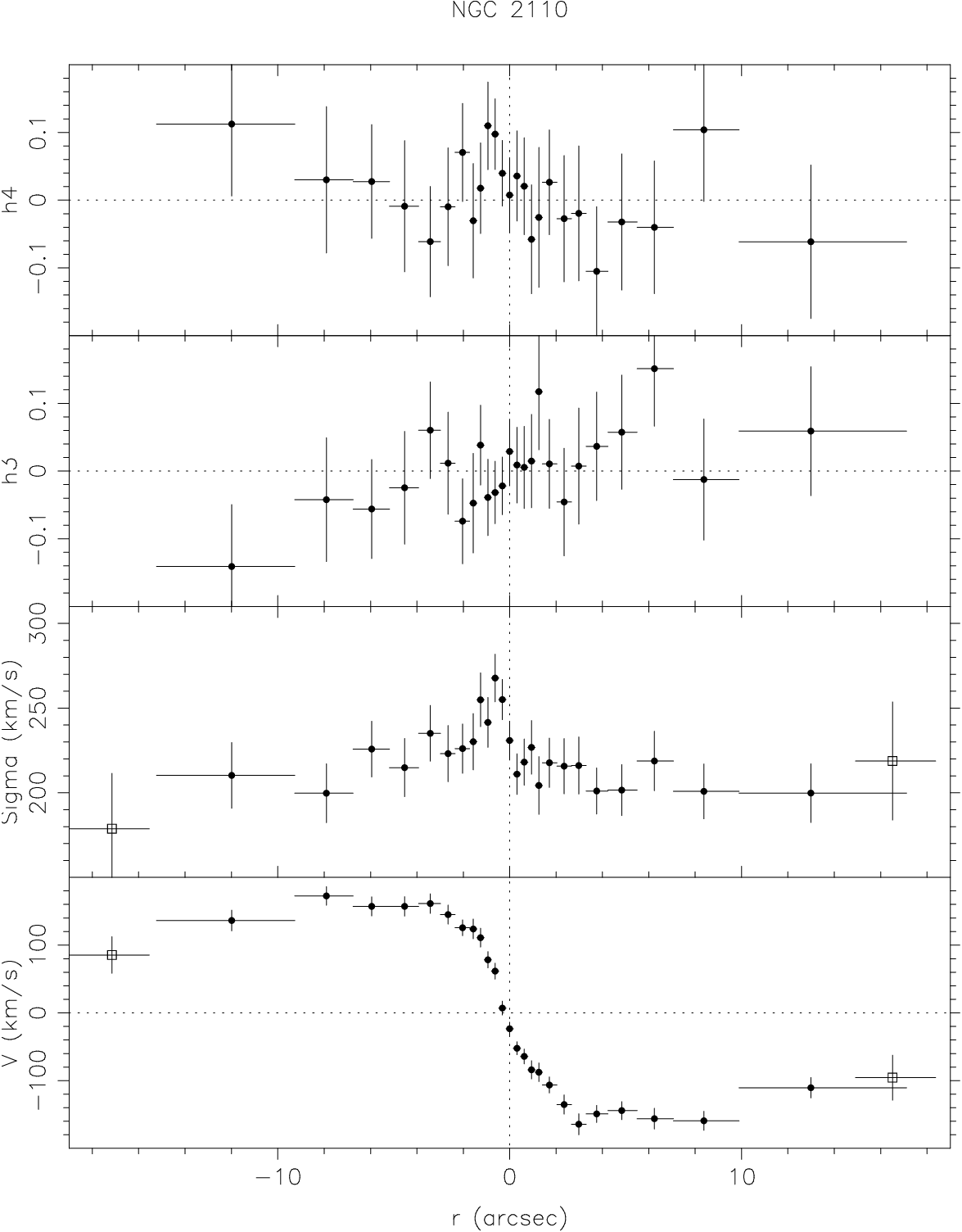}
\includegraphics[scale=0.42]{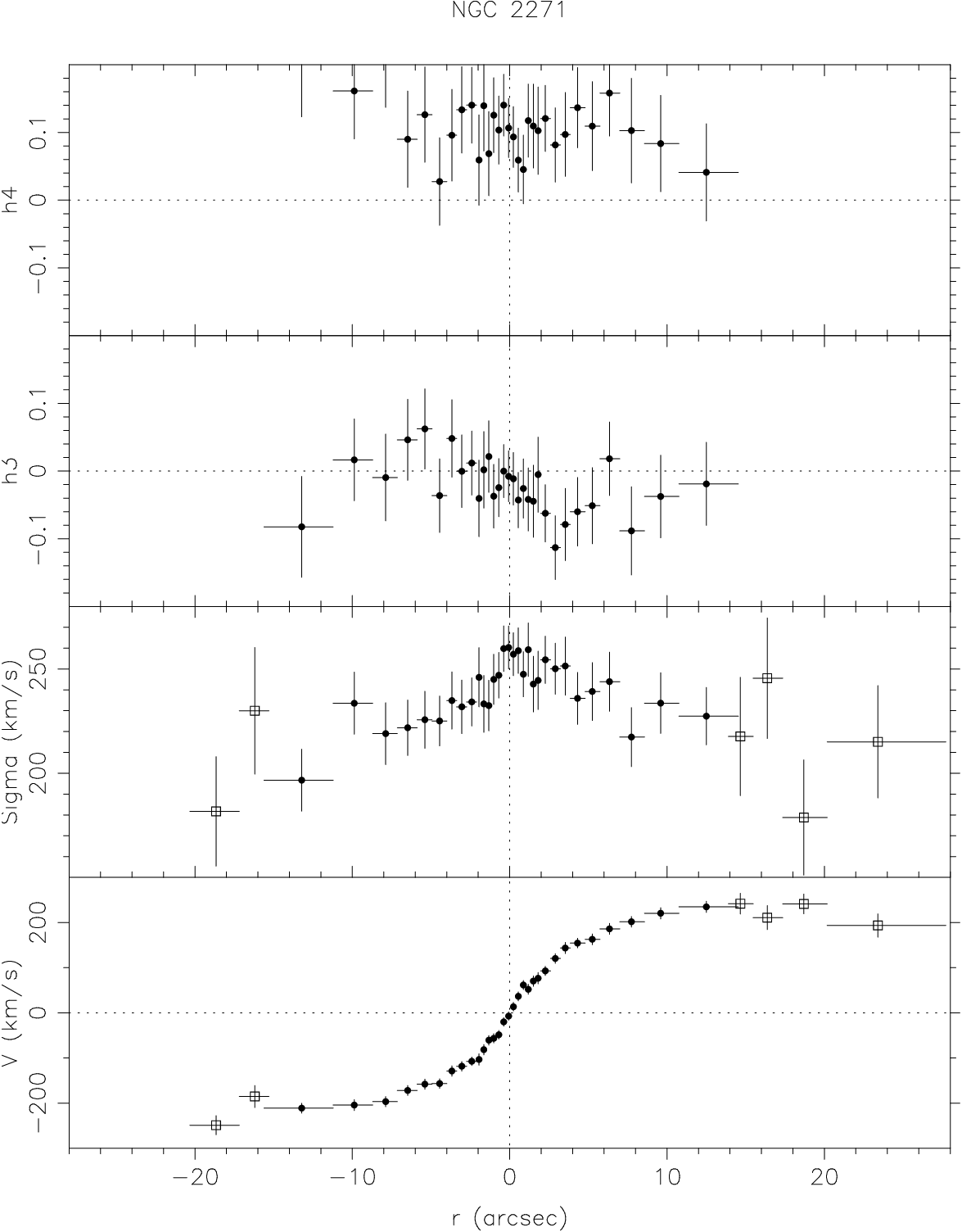}
\caption{Radial kinematic profiles of isolated galaxies. From
bottom to top are shown rotation velocity, velocity dispersion,
3rd and 4th hermite terms. The open symbols indicate lower S/N
data for which we do not calculate hermite terms.}
\end{center}
\end{figure*}

\begin{figure*}
\begin{center}
\includegraphics[scale=0.42]{N2865vsh3h4.ps}
\includegraphics[scale=0.42]{N4240vsh3h4.ps}
\includegraphics[scale=0.42]{M0127013vsh3h4.ps}
\includegraphics[scale=0.42]{M0213009vsh3h4.ps}
\caption{Radial kinematic profiles of isolated galaxies. From
bottom to top are shown rotation velocity, velocity dispersion,
3rd and 4th hermite terms. The open symbols indicate lower S/N
data for which we do not calculate hermite terms.}
\end{center}
\end{figure*}

\begin{figure*}
\begin{center}
\includegraphics[scale=0.42]{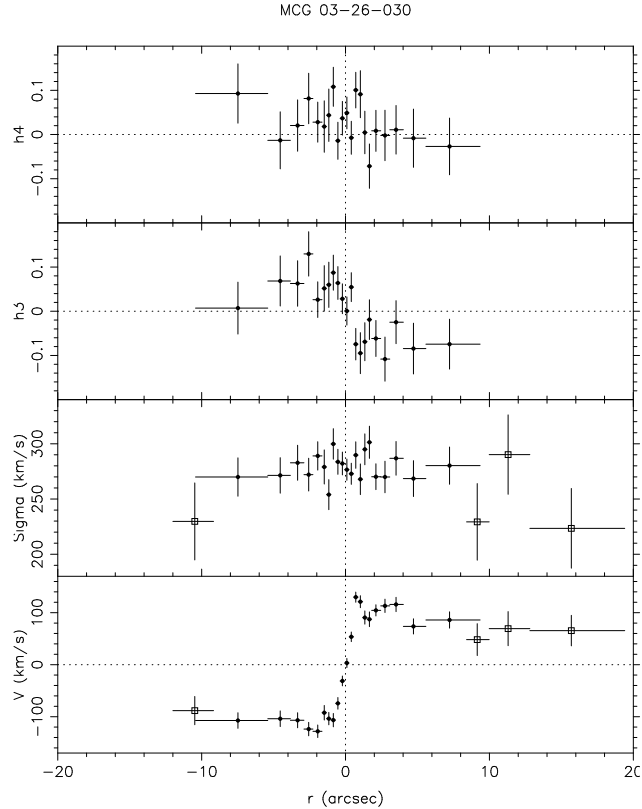}
\caption{Radial kinematic profiles of MCG-03-26-030. From
bottom to top are shown rotation velocity, velocity dispersion,
3rd and 4th hermite terms. The open symbols indicate lower S/N
data for which we do not calculate hermite terms.}
\end{center}
\end{figure*}

\begin{table*}
\begin{center}
\begin{tabular}{lccccc}
\hline
Galaxy & $V_{system}$ & $V_{max}$ & $\sigma_0$ &
$(V/\sigma_0)^{\ast}$ & $\sigma_0$ (lit)\\
& (km/s) & (km/s) & (km/s) & & (km/s)\\
\hline
ESO 153-G003 & 	6341.5	& $42.7	\pm 18.8$ & $222.5 \pm 11.9$ &
$0.19 \pm 0.09$ & --\\
ESO 218-G002  & 4251.3 & $44.0 \pm   7.4$ & $297.9 \pm 11.2$ &
$0.30 \pm 0.06$ & 235\\
ESO 318-G021  & 4793.3 & $152.9 \pm 16.5$ & $256.5 \pm 8.9$ &
$0.92 \pm 0.13$ & 156\\
NGC 682 &	5436.8	& $74.5 \pm 12.6$	& $199.4 \pm 4.7$
& $0.58 \pm 0.11$ & --\\
NGC 1045 &	4462.4	& $106.4 \pm 17.6$	& $286.1 \pm 5.1$
& $0.57 \pm 0.10$ & --\\
NGC 1162 &	3752.3	& $103.9 \pm 4.7$	& $208.3 \pm 5.3$
& $1.52 \pm 0.11$ & --\\
NGC 2110      & 2326.3 & $168.5 \pm 10.3$ & $230.8 \pm 11.4$ &
$1.46 \pm 0.16$ & 246\\
NGC 2271      & 2560.6 & $245.2 \pm 11.5$ & $260.3 \pm 10.3$ &
$1.45 \pm 0.12$ & 148\\
NGC 2865      & 2558.1 & $109.7 \pm 7.7$  & $183.1 \pm 5.3$ &
$1.20 \pm 0.12$ & 170\\
NGC 4240      & 1878.6 & $83.6 \pm 12.3$  & $129.4 \pm 6.4$ &
$1.96 \pm 0.38$ & --\\
MCG-01-27-013 & 9051.5 & $38.9 \pm 14.0$  & $250.9 \pm 9.8$ &
$0.17 \pm 0.07$ & --\\
MCG-02-13-009 & 5533.9 & $25.0 \pm 14.1$  & $245.2 \pm 9.2$ &
$0.31 \pm 0.19$ & --\\
MCG-03-26-030 & 8925.0 & $129.0 \pm 7.8$  & $276.6 \pm 9.6$ &
$0.80 \pm 0.08$ & --\\
\hline
\end{tabular}
\caption{Kinematic measurements.  
  The systemic velocity $V_{system}$ is taken as the mean of the
  velocities measured for the spectra with $S/N \ge 30 $. The maximum
  rotational velocity $V_{max}$ is taken as the difference of the most
  extreme velocity measurements for all data points.  The central
  velocity dispersion $\sigma_0$ is taken as the measurement for the spatial bin
  closest to the galaxy centre. The anisotropy parameter
$(V/\sigma_0)^{\ast}$ is defined as 
$(V_{max}/\sigma_0) / \sqrt{\epsilon/1 - \epsilon}$. Literature
velocity dispersions come from Hyperleda. 
}
\label{'vsystem'}
\end{center}
\end{table*}

\begin{table}
\begin{center}
\begin{tabular}{lccc}
\hline
Galaxy & $\sigma_{{\rm norm}}$ & $S/N$   \\
       & (km/s)                &          \\
\hline
ESO 153-G003 & $223.5 \pm 2.9$ &170\\
ESO 218-G002 & $284.1 \pm 3.3$ &136\\
ESO 318-G021 & $244.4 \pm 3.0$ &148\\
NGC 682      & $203.9 \pm 1.8$ & 235\\
NGC 1045     & $269.2 \pm 1.8$ & 260\\
NGC 1162     & $209.1 \pm 1.8$ & 235\\
NGC 2110     & $253.8 \pm 4.0$ &145\\
NGC 2271     & $248.6 \pm 3.0$ &142\\
NGC 2865     & $176.8 \pm 1.8$ &288\\
NGC 4240     & $120.4 \pm 2.3$ &194\\
MCG-01-27-013 & $246.6 \pm 3.5$ &123\\
MCG-02-13-009 & $233.9 \pm 3.5$ &125\\
MCG-03-26-030 & $318.1 \pm 3.6$ &152\\
\hline
\end{tabular}
\caption{Central velocity dispersion measurements. Values are derived
  from spectra extracted over the central $5\asec$ along the slit to
  match a standard aperture of 3.4\asec for a galaxy at the distance
  of Coma.  $\sigma_{norm}$ is the dispersion extracted from a fixed
  aperture of $5.0\asec \times 1.5\asec$ or $5.0\asec \times 1.2\asec$
  and normalised using the aperture prescription
  of Jorgensen et al. (1995). 
$S/N$ is the signal-to-noise ratio
  of the extracted spectrum.  }
\label{'vsystem'}
\end{center}
\end{table}

\subsection{ESO 153-G003}

Classified as an S0 galaxy in NED, ESO 153-G003 reveals a thin
dust lane along the major axis and may actually be an edge-on Sa
galaxy. 

We placed the slit along the minor axis for this galaxy. In this
case, the velocity profile is generally consistent with no
rotation in the outer regions. The inner regions suggest some
velocity substructure but this could be due to a small
misalignment of the slit by 1 arcsec or the major axis dust
lane. The velocity dispersion profile is fairly flat within the
inner region indicative of a rotating disk. The \h3 and \h4 terms
are inconclusive.

\subsection{ESO 218-G002}

From the imaging study of Reda et al. (2004), ESO 218-G002 has a
large position angle twist in its isophotes but otherwise shows
no obvious fine structure. ESO 218-G002 lies in the direction of
the `Great Attractor' (Dressler, Faber \& Burstein 1991), 
and near the border
of the so-called Norma cluster.

Along the slit
at radii $\approx 15^{''}$ on both sides of nucleus there is
an excess of light due to faint sources which are too faint for redshift
measurements. There is a faint, marginally resolved source with
an emission line spectrum at 21.2$^{''}$ from the galaxy nucleus. Its
position measured from a b$_J$ image of the SuperCosmos Sky Survey is
$\alpha$ = 12 21 08.26, $\delta$ = --52 35 30.8 (J2000). Emission lines
of $H\gamma$, $H\delta$, $H\beta$ and [OIII] are present
and the redshift measured from these is
$z= 0.101$. Close to this source at $\alpha$ = 12 21 08.44 $\delta$ =
--52 35 35.3 is another marginally resolved source which has no
emission in its spectrum and we are unable to measure a redshift from
its faint absorption-line spectrum.
It may be possible that there is a background cluster in the
vicinity of ESO 218-G002. 
At 83 arcsec from ESO 218-G002 there is an X-ray source (1RXS
J122104.3-523617).

The velocity profile of ESO 218-G002 is roughly solid-body but lop-sided. The
rotation is slow, peaking at 44 km/s. The velocity dispersion
rises towards the centre with a corresponding dip in the central \h3
profile. 

\subsection{ESO 318-G021}

The galaxy shows no obvious fine structure (Reda et al. 2004).  There
is weak [OIII] emission at the centre of the galaxy which is also
slightly extended. The gas rotation is the same sense and same
magnitude as the stars. There may also be some H$\beta$ emission but
it is weak and rotation is not obvious. There is strong \hbeta\ 
absorption indicative of a young nuclear stellar population.

We find a smooth regular velocity profile, and declining velocity
dispersion with radius. These properties
are consistent with a fast rotator containing a disk.
Within the errors both the \h3 and \h4
terms show relatively flat profiles. 

\subsection{NGC 682}

The DSS and 2MASS imaging suggests a featureless elliptical.  The
velocity profile is lop-sided suggesting a possible central KDC of
size 4 arcsec (1.4 kpc). In the same central regions the velocity
dispersion profile is relatively flat. The hermite terms suggest a
non-equilibrium structure. We note that the spectrum was taken at an
angle intermediate to the major and minor axes.


\subsection{NGC 1045}

NGC 1045 reveals a tidal plume, evidence for dust and boxy
isophotes (Reda et al. 2004), suggestive of a recent merger. 
The velocity profile shows solid-body rotation rising to
$\sim$100 km/s at our largest radii of $\sim$25
arcsec (7.3 kpc). The velocity dispersion is sharply peaked at the galaxy
centre. The  \h3 and \h4 profiles (and the inner rotation profile) 
suggest some velocity substructure.

\subsection{NGC 1162}

The DSS and 2MASS imaging suggest a featureless elliptical, with
fairly round inner regions. 
The galaxy major axis is therefore difficult to define. We
estimate a position angle
(P.A.) of around 51 degrees, whereas we placed our slit on an
intermediate axis of P.A. = 280 degrees.   
The central regions show evidence for a rotating disk. 
The velocity dispersion profile reveals a dip in the central 3 arcsec. 

\subsection{NGC 2110}

The galaxy is a well known AGN (e.g. Pfefferkorn et al. 2001) and
appears in the isolated galaxy sample of Colbert et al. (2001).
Strong emission lines are seen, with a large
spatial extension beyond the continuum. This may affect the
kinematic values derived in the outer, lower S/N, region.
Strong rotation is seen with a (off-centre) peaked velocity
dispersion profile. Both hermite profiles  appear to be fairly
regular. 

\subsection{NGC 2271}

The DSS and 2MASS imaging indicates an elongated ($\epsilon$ =
0.3) early-type galaxy. 
Koprolin \& Zeilinger (2000) have suggested that NGC 2271
contains a KDC within the central 1.5$^{''}$ although it is not
obvious from their kinematic profiles. They measure
a rotation between the flat part of the rotation curve and the
central regions of 263 $\pm$ 36 km/s and a velocity dispersion
within the central 5 arcsec of $\sigma_o$ = 237
$\pm$ 7 km/s. The (V/$\sigma$)$^{\ast}$ = 1.72 $\pm$ 0.24, which
suggests an anisotropic rotator. We measure a maximum velocity of
245 $\pm$ 12 km/s and $\sigma_o$ = 260 $\pm$ 10 km/s. The
resulting (V/$\sigma$)$^{\ast}$ = 1.44 $\pm$ 0.12 (see Table 3). 

We find a smooth regular velocity profile, and declining velocity
dispersion. The rotation reaches a high value of 245 km/s. The \h3
profile generally supports the presence of a disk. We are unable
to confirm the claim of Koprolin \& Zeilinger (2000) of a KDC in
NGC 2271. 

\subsection{NGC 2865}

This galaxy contains dust and shells (Reda et al. 2004) and is an
excellent candidate for a merger remnant (Hau et al. 1999). 
Koprolin \& Zeilinger (2000) measure little or no rotation of 72
$\pm$ 15 km/s, $\sigma_o$ = 184 $\pm$ 5 km/s, and
(V/$\sigma$)$^{\ast}$ = 0.67 $\pm$ 0.14. We measure a maximum
velocity of 110 $\pm$ 8 km/s and $\sigma_o$ = 183 $\pm$ 5
km/s. Combined with an assumed ellipticity of 0.2 gives 
(V/$\sigma$)$^{\ast}$ = 1.20 $\pm$ 0.12 (see Table 3). 

The velocity profile shows a clear signature of a KDC in the
central 4 arcsec (6.4 kpc) and solid-body rotation in the outer
regions. Within the same central region, the
velocity dispersion reveals a strong dip from $\sim$ 200 to 160
km/s. The \h3 and \h4 profiles are quite distorted indicating a
non-equilibrium situation. 

\subsection{NGC 4240}

The DSS imaging suggests a featureless, nearly round elliptical with a
nearby bright star to the West.  The galaxy reveals a KDC in the
central 4 arcsec (5.0 kpc) and a lop-sided rotation (i.e. 100 km/s vs
50 km/s). The velocity dispersion profile is quite flat. The \h3 and
\h4 profiles peak at the galaxy centre.

\subsection{MCG-01-27-013}

The DSS and 2MASS imaging suggest a featureless elliptical. 
The velocity profile reveals evidence for a weak KDC within 5
arcsec (2.9 kpc) and little
or no rotation in the outer regions. The velocity dispersion
profile is fairly constant with radius. The hermite profiles 
indicate a non-equilibrium structure. 

\subsection{MCG-02-13-009} 

The DSS imaging indicates several faint galaxies, but redshifts are
not available for them. The galaxy 
has E+A type spectra and [OIII] emission.
The galaxy shows some evidence of slow rotation in the inner regions
but is consistent with no rotation in the outer parts. A
centrally peaked velocity dispersion profile is seen. 

\subsection{MCG-03-26-030}

The imaging study of Reda et al. (2004) found this galaxy to have
disky isophotes. 
The velocity profile shows evidence for rapid rotation in the
inner 4 arcsec (2.3 kpc), reaching a constant velocity of almost 100 km/s
in the outer regions. The velocity dispersion is fairly
constant. The \h3 term shows an symmetric profile 
indicating a disky component is present.

\begin{figure}
\begin{center}
\includegraphics[scale=0.4,angle=0]{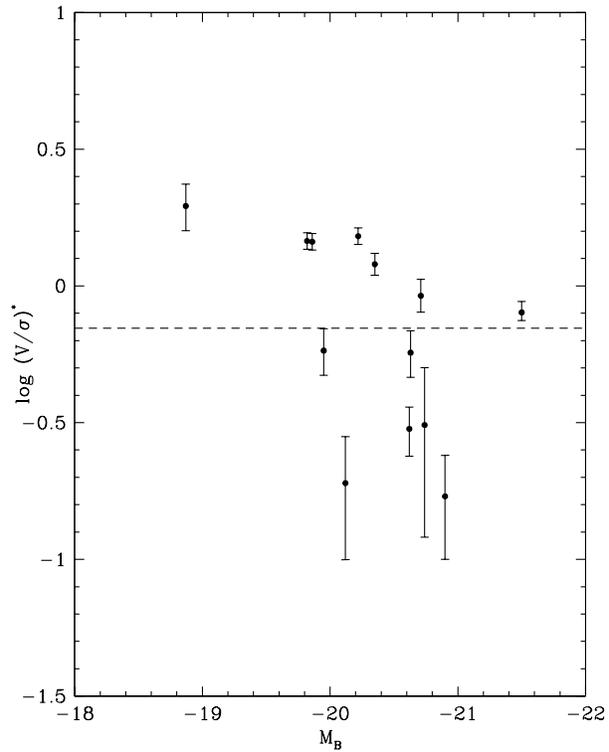}
\caption{Anisotropy-luminosity diagram. Anisotropy parameter as a
function of galaxy B-band luminosity for isolated galaxies. The
horizontal dashed line shows the separation between rotation
supported galaxies and anisotropic ones as given by Bender et
al. (1992). Our small sample is consistent with the general trends for
galaxies in higher density environments. }
\end{center}
\end{figure}

\begin{figure}
\begin{center}
\includegraphics[scale=0.4,angle=0]{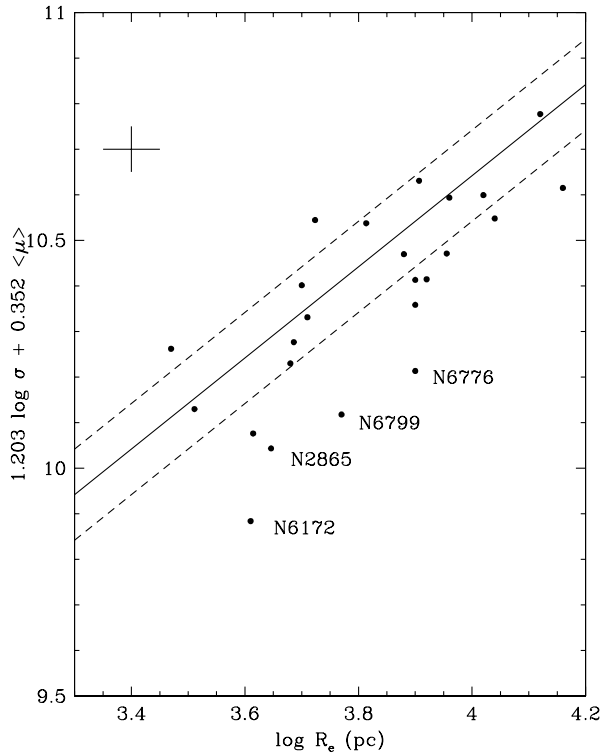}
\caption{Fundamental Plane for isolated galaxies. Isolated galaxies
  shown are from this work and from Reda, Forbes \& Hau (2005). The
  solid and dashed lines show the Coma cluster relation and dispersion
  from Jorgensen \etal (1993). Isolated galaxies reveal a similar FP to
  those of cluster ellipticals, albeit with somewhat larger
  scatter. Four particularly deviant galaxies are labelled (see text
  for details).  }
\end{center}
\end{figure}

\section{Discussion}
The internal kinematics of early-type galaxies has been investigated
by several authors in the past (e.g. Bender, Saglia \& Gerhard 1994;
Franx, Illingworth \& Heckman 1989; Mehlert \etal 1997; Koprolin \&
Zeilinger 2000). Recently this has included 2D kinematics via integral
field units, e.g. Cappellari \etal (2005). Most of these previous
studies covered early-type galaxies in clusters and groups -- none
focused on isolated galaxies. 
KDCs are thought to be small, rapidly
rotating disks at the centres of these galaxies, although they are
very difficult to detect photometrically (Forbes, Franx \& Illingworth
1995).

Excluding our one minor-axis observation, the remaining dozen
galaxies reveal evidence for a KDC or velocity substructure in 5
cases (an additional KDC has been claimed in NGC 2271 but it is
not seen in our data). Thus KDCs are fairly common in isolated
galaxies at around 40\%. 

From these studies of galaxies in high density environments, 
the incidence of KDCs is observed to be about
33\% (rising to 50\% when projection effects are
considered). Thus we find a similar, possibly even higher,
fraction in isolated galaxies. Koprolin \& Keilinger
(2000) found typical KDC radii of 0.4 kpc. For our sample, which is
on average 2-3 times more distant than the Koprolin \&
Keilinger sample, we find the KDCs to have diameter sizes of a
few kpc. KDCs in galaxies located in high density environments 
are generally found to be old and hence long-lived systems
(Kuntschner et al. 2005). We note that the KDC in NGC
2865 with an age of $\le$ 1 Gyr (Hau et al. 1999) is an
exception. A future paper will explore radial stellar population
trends, and derive KDC ages, for this sample of 13 galaxies. KDCs are generally thought
to be the result of a major merger (Hernquist \& Barnes 1991),
however they may also form in an early collapse without
subsequent mergers (Harsoula \& Voglis 1998).

The anisotropy parameter $(V/\sigma)^{\ast}$ can be used to separate
galaxies that are rotationally supported (generally low luminosity)
from those that are anisotropic (generally high luminosity). The
division occurs around $(V/\sigma)^{\ast}$ = 0.7 (Bender et
al. 1992). Our sample of 13 galaxies follow the trends seen for larger
samples (e.g. Davies et al. 1983; Bender et al. 1992; Koprolin \&
Zeilinger 2000). Although as Fig. 5 shows, our sample reveals a
wide range of anisotropy values for a small range in galaxy
magnitude. This general trend has been
interpreted as a change in the merger progenitor type from
gaseous (wet) mergers 
of spirals to the gas-free (dry) mergers of ellipticals (Naab,
Khochfar \& Burkert 2005). 
Perhaps the only notable exception is MCG-03-26-030
which appears to be a high luminosity galaxy (M$_B$ = --21.5) that is
rotationally supported. There is evidence for a disk in the kinematic
profile of this galaxy (see Fig. 4).

As noted in the Introduction, Weil \& Hernquist (1996) have
predicted that remnants of multiple major mergers (e.g. a merged
small group) will reveal rotation about the minor axis and an increasing
V/$\sigma$ with galactocentric radius. In virtually all cases
(the exception being ESO153-G003 where we placed the slit along
the minor axis) our isolated galaxy sample reveals both of these
properties. However, we note that such properties are also
commonly seen in ellipticals in higher density environments
(e.g. Koprolin \& Zeilinger 2000). The strongest kinematic
differences between pair and multiple mergers are predicted to
arise at 3--4 effective radii, i.e. well beyond the radii that
current long-slit observations can explore. Perhaps the best argument
against our sample being the remnants of merged groups is their
modest optical luminosity. Only MCG-03-26-030 with M$_B$ = --21.5
is bright enough to be a possible small `fossil' group (see
Jones, Ponman \& Forbes 2000).

In Fig. 6 we show the FP for isolated galaxies compared to Coma
cluster ellipticals (Jorgensen \etal 1993). For the isolated
galaxies we include the galaxies studied in this work plus those
from Reda, Forbes \& Hau (2005) where available. The plot shows
that isolated galaxies generally follow the FP of ellipticals in
the higher density environment of the Coma cluster with a slightly
larger scatter ($\sim$0.15 dex compared to 0.1 dex) and offset by
-0.1 dex. Four galaxies in particular are strongly offset
from the mean FP (i.e. NGC 2865, NGC 6172, NGC 6776 and NGC 6799). This
offset can be understood in terms of their young stellar
populations and hence reduced mass-to-light  ratio (Forbes,
Ponman \& Brown 1998). The offset in the remaining galaxies could
be a stellar population effect or a small calibration offset. 

\section{Conclusions}

We have derived rotation velocity, velocity dispersion and hermite
radial profiles for 13 isolated early-type galaxies. These
profiles extend to $\sim$ 1 effective radius, or half of the
enclosed galaxy mass. We 
observe a dichotomy in kinematic properties similar to that in the
elliptical population as a whole, where rotational support
tends to be found in low luminosity ellipticals. For all galaxies
the V/$\sigma$ ratio increases with radius. 
We find kinematically distinct cores (KDCs) or velocity
substructure in $\sim 40\%$ of the galaxies for which we have
major axis spectra. Such a fraction is similar to that observed
for ellipticals in higher density environments.  
Most galaxies in the sample reveal kinematic evidence for a
nuclear disk. The non-relaxed
kinematics in several galaxies suggests that they have undergone 
mergers or accretion events. 
Isolated ellipticals generally follow the
fundamental plane defined by cluster ellipticals -- exceptions
being those galaxies with evidence for young stellar
populations. Overall, isolated ellipticals have similar kinematic
properties to their counterparts in higher density environments. 
Our conclusions support those from our previous papers on
isolated galaxies, i.e. Reda et al. (2004, 2005).

\section*{Acknowledgments}

We thank R. van der Marel for his spectral analysis code. We also
thank R. Proctor and S. Brough for their help carrying out some
of the observations. We thank F. Reda and S. Brough for useful comments on the
text. DF thanks the ARC for partial financial support.



\end{document}